\documentclass[twocolumn,showpacs,preprintnumbers,amsmath,amssymb,superscriptaddress]{revtex4}

\usepackage{calc}
\usepackage{float}
\usepackage{graphicx}
\usepackage{bm}

\begin{document}

\title{Non-Gaussianity and Dynamical Trapping in Locally Activated Random Walks}

\author{O. B\'enichou}
\affiliation{Laboratoire de Physique Th\'eorique de la Mati\`ere Condens\'ee, CNRS UMR 7600, case courrier 121, Universit\'e Paris 6, 4 Place Jussieu, 75255
Paris Cedex}

\author{N. Meunier}
\affiliation{MAP5, CNRS UMR 8145, Universit\'e Paris Descartes, 45 rue des Saints-P\`eres, 75270 Paris Cedex 06, France} 

\author{S. Redner}
\affiliation{Center for Polymer Studies and Department of Physics,
Boston University, Boston, Massachusetts 02215, USA}

\author{R. Voituriez}
\affiliation{Laboratoire de Physique Th\'eorique de la Mati\`ere Condens\'ee, CNRS UMR 7600, case courrier 121, Universit\'e Paris 6, 4 Place Jussieu, 75255
Paris Cedex}

\date{\today}

\begin{abstract}

  We propose a minimal model of \emph{locally-activated diffusion}, in which
  the diffusion coefficient of a one-dimensional Brownian particle is
  modified in a prescribed way --- either increased or decreased --- upon
  each crossing of the origin.  Such a local mobility decrease arises in the
  formation of atherosclerotic plaques due to diffusing macrophage cells
  accumulating lipid particles.  We show that spatially localized mobility
  perturbations have remarkable consequences on diffusion at all scales, such
  as the emergence of a non-Gaussian multi-peaked probability distribution
  and a dynamical transition to an absorbing static state.  In the context of
  atherosclerosis, this dynamical transition can be viewed as a minimal
  mechanism that causes macrophages to aggregate in lipid-enriched regions
  and thereby to the formation of atherosclerotic plaques.
\end{abstract}


\maketitle

Many-particle systems that consume energy for self-propulsion --- active
particle systems --- have received growing attention in the last decade, both
because of the new physical phenomena that they display and their wide range
of applications.  Examples include molecular motors, cell assemblies, and
even larger organisms \cite{Toner:2005}. The intrinsic out-of-equilibrium
nature of these systems leads to remarkable effects such as non-Boltzmann
distributions \cite{PhysRevE.59.5582}, long-range order even in low spatial
dimensions \cite{PhysRevLett.75.4326} and spontaneous flows
\cite{Voituriez:2005}.

At the single-particle level, the active forcing of a Brownian particle leads
to non-trivial statistics.  For example, it has been recently shown
\cite{PhysRevLett.106.160601,2011arXiv1107.4225E} that a random walk which is
reset to its starting point at a fixed rate has a non-equilibrium stationary
state, as opposed to standard Brownian motion.  Another example is given by
self-propelled Brownian particles \cite{PhysRevLett.106.230601}, which can
yield sharply peaked probability densities for the particle velocity.

\begin{figure}[ht]
\begin{center}
\includegraphics[width=0.5\textwidth]{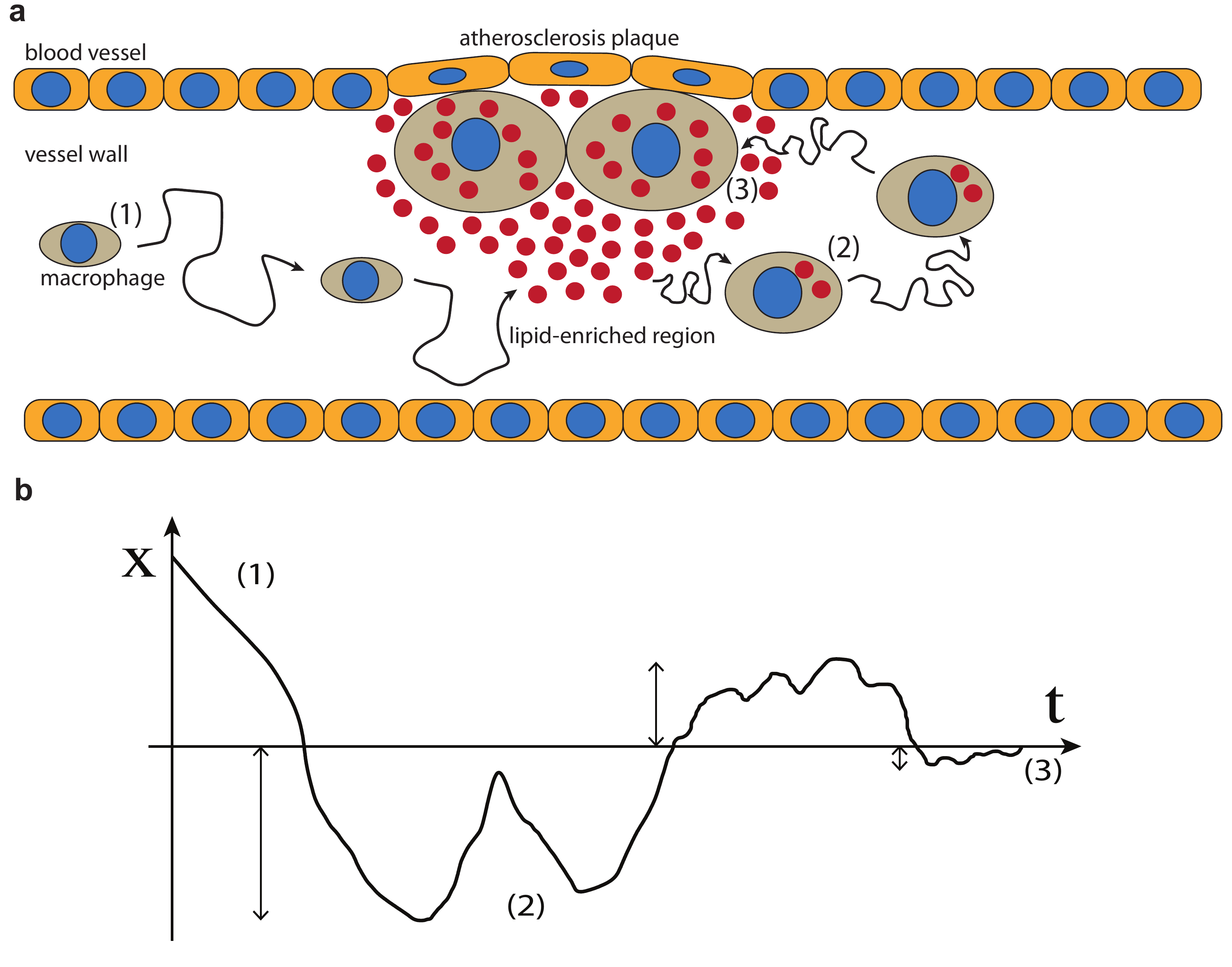}
\caption{{\bf a.} Sketch of the different stages of atherosclerosis plaque
  formation: (1) rapid diffusion of a ``free'' macrophage cell; (2) upon
  entering a localized lipid-enriched region, the macrophage accumulates
  lipids, and thereby grows and becomes less mobile; (3) after many crossings
  of the lipid-enriched region, the macrophage eventually gets trapped,
  resulting in the formation of an atherosclerotic plaque.  {\bf b.} Sketch
  of a one-dimensional particle trajectory of the model of locally
  decelerated random walk. }
\label{plaque}
\end{center}
\end{figure}


In this letter, we consider a new class of problems in which the active
forcing of a Brownian particle is {\it localized in space}.  While the impact
of localized perturbations on random walks has been
investigated~\cite{Hughes:1995a}, in part because of its relevance to a wide
range of situations, such as localized sources and
sinks~\cite{PhysRevA.30.3362,SN1990207}, trapping
\cite{Weiss:1994,Tejedor:2010a} or diffusion with forbidden \cite{L.:1960},
hop-over \cite{Zia:1998} or defective \cite{Benichou:2002qq} sites, the role
of local activation on Brownian-particle dynamics remains open.  We present a
minimal model of locally activated diffusion, in which the diffusivity of a
Brownian particle is modified --- either increased or decreased --- in a
prescribed way upon each crossing of the origin.  

A prototypical example is a bacterium in the presence of a localized patch of
nutrients, which enhances the ability of the bacterium to move, or,
alternatively, toxins that impair bacterial mobility.  This type of localized
decrease of mobility also underlies the dynamics of a cell (e.g., a
macrophage) that grows by accumulating smaller and spatially localized
particles, such as lipids (Fig.~\ref{plaque}) \cite{Calvez:2008,Calvez:2009}.  As the cell grows, its
ability to move decreases and the ultimate result is the formation of an
atherosclerotic plaque \cite{Winther:2000}.  The spatial localization arises
from the presence of lipids at specific points in the arterial network; these
lipids can be located, as is now well accepted, by the properties of the
blood flow \cite{Caro:2009}.  Observations show that
macrophages that have accumulated lipids move more slowly.  Eventually the
macrophage stops in an lipid-enriched region, resulting in the formation of
an atherosclerotic plaque \cite{Yvan-Charvet:2007,Siegel-Axel:2008}.  Here we
propose a simple model to account for this local mobility decrease and
address the particular questions of (i) the potential trapping of cells in
locally lipid-enriched regions, and (ii) the kinetics of the resulting
segregation process when it exists.

Our formalism allows us to describe both the situations of decreased and
increased localized mobility changes.  We show that this type of perturbation
has remarkable consequences on the diffusion process at all scales.  We
stress that the diffusion coefficient of the active particle at any time
depends on the entire history of the trajectory.  Thus the evolution of the
particle position is intrinsically non-Markovian
\cite{Kampen:1992a,PhysRevLett.77.1420,Majumdar:1999a,Redner:2001a}.  Our
main findings are: (i) The probability distribution of the position has a
non-Gaussian tail.  (ii) For local acceleration, a diffusing particle is
repelled from the origin, so that the maximum in the probability distribution
is at non-zero displacement.  (iii) For local deceleration, a dynamical
transition to an absorbing state occurs.  For sufficiently strong
deceleration, the particle can get trapped at the origin at a finite time.
The exact time dependence for the particle survival probability is determined
explicitly.  Conversely, if the deceleration process is sufficiently weak,
the particle never gets trapped.  This dynamical transition to an absorbing
state provides a minimal mechanism that could help understand the formation
kinetics of atherosclerotic plaques.

{\it The Model.}  A one-dimensional diffusing particle is accelerated or
decelerated whenever it crosses the origin $x=0$ according to the following
Langevin equations:
\begin{equation}
\label{LE}
\dot{x}=\sqrt{2D}\,\xi(t)\,,\qquad\qquad
\dot{D}= f(D) \,\delta(x)\,,
\end{equation}
where $\xi$ is a Gaussian white noise of intensity one, $D$ the particle
diffusion coefficient, $x$ the particle position, $\delta(x)$ the Dirac
distribution, and $f(D)$ an arbitrary prescribed function that accounts for
the local activation.  For simplicity, we assume that the particle is
initially at $x=0$ with $D=D_0>0$.  Note that: (i) both the position $x$ and
the diffusion coefficient $D$ are random variables; (ii) as mentioned
previously, the evolutions of $x$ or $D$ alone are non-Markovian; (iii) the
function $f(D)$ can be positive (local acceleration) or negative (local
deceleration), but with $f(0)=0$ so that $D$ remains non negative.

Following standard steps, the corresponding Fokker-Planck equation
\cite{Gardiner:2004} for the joint distribution of position $x$ and diffusion
coefficient $D$ at time $t$, $P(x,D,t)$, is:
\begin{align}
\label{FP}
\frac{\partial P}{\partial t}=D\frac{\partial^2 P}{\partial x^2}
-\delta(x)\frac{\partial [f(D) P]}{\partial D}
-\lambda(t) \delta(x)\delta(D),
\end{align}
where the last term of the right side accounts for the absorbing state at
$(x=0,D=0)$.  The explicit expression for $\lambda(t)$ is determined
demanding that $P$ is normalized, from which we obtain
\begin{equation}
\label{def lambda}
\lambda(t)=\lim_{D\to0}\big[f(D)P(0,D,t)\big].
\end{equation}
When $f(D)$ is positive, then $D$ is always non zero.  In this case, the
particle is never trapped and $\lambda(t)=0$ at all times.  While intuitively
obvious for local acceleration ($f(D)>0$), we show below that $\lambda(t)$
can equal zero for local deceleration processes.

{\it Local Acceleration, $f(D)\!>\!0$.}  Laplace transforming Eq.~(\ref{FP}) gives
\begin{equation}
\label{LT}
-s{\widehat P+ D\frac{\partial^2 {\widehat P}}{\partial x^2}}= 
\delta(x)\Big[\frac{\partial (f{\widehat P})}{\partial D}-\delta(D-D_0)\Big]\,,
\end{equation}
where $\widehat{P}=\widehat{P}(x,D,s)$ is the Laplace transform of the
probability distribution.  For $x\neq0$, the solution is
\begin{equation}
\label{morceaux}
{\widehat P}(x,D,s)=A(D,s)\,e^{-|x|\,\sqrt{{s}/{D}}}~,
\end{equation}
where the coefficient $A(D,s)$ is determined by integrating Eq.~\eqref{LT}
across $x=0$ to obtain the jump of the first derivative of ${\widehat P}$
with respect to $x$ at this point:
\begin{align*}
D\left[\frac{\partial {\widehat P}}{\partial x}\Big|_{x=0+}\! -\frac{\partial{\widehat P}}{\partial x}\Big|_{x=0-}\! \right]=
\frac{\partial [f{\widehat P}(x\!=\!0)]}{\partial D}-\delta(D\!-\!D_0)\,.
\end{align*}
Using Eq.~(\ref{morceaux}), we have
\begin{equation}
\label{fD}
f\frac{\partial A}{\partial D}+\left[f'+2\sqrt{sD}\right] A=\delta(D-D_0).
\end{equation}
When $f(D)$ is positive, then $A(D,s)=0$ for $D<D_0$, while for $D>D_0$ the
solution to \eqref{fD} is
\begin{equation}
  A=B(s)\,\frac{f(D_0)}{f(D)}\,e^{-\sqrt{4s}\,F}~,
\end{equation}
with 
\begin{equation*}
F(D)\equiv \int_{D_0}^D \frac{\sqrt{D'}}{|f(D')|}\,{\rm d}D'\,.
\end{equation*}
The unknown function $B(s)$ is determined by the jump of $A$ at $D_0$:
\begin{equation*}
A(D_0^+,s)-A(D_0^-,s)=\frac{1}{f(D_0)}~,
\end{equation*}
which finally yields
\begin{equation}
{\widehat P}(x,D,s)=\Theta(D\!-\!D_0)\frac{1}{f(D)}\,e^{-Z\sqrt{s}}~,
\end{equation}
where $\Theta$ is the Heaviside step function and
\begin{equation*}
Z\equiv\frac{|x|}{\sqrt{D}}+2F(D)\,.
\end{equation*}
Laplace inverting this expression, we obtain the joint distribution
\begin{equation}
\label{joint}
P(x,D,t)=\Theta(D-D_0)\frac{Z}{f(D)\sqrt{4\pi t^3}}\,\, e^{-Z^2/4t}~.
\end{equation}

The marginal distribution with respect to $x$, that is, the probability
distribution of positions, is obtained by integrating Eq.~\eqref{joint} over
all $D$ in the range $[D_0,\infty]$.  While it does not seem possible to
evaluate this integral analytically, the large-$x$ behavior can be obtained
by the Laplace method.  For the illustrative case where $f(D)$ is a constant
(that we define as $a$), this method gives
\begin{equation}
\label{large x}
P(x,t)\sim
\frac{1}{t}\sqrt{\frac{|x|}{3a}}\,\exp\left[-\frac{8|x|^{3/2}}{9\sqrt{a}\,t}\right]\qquad
x\to\infty\,,
\end{equation}
which we numerically checked is close to the exact value $P(x,t)$ over a wide
spatial range.  We wish to emphasize two important features of this result
for $P(x,t)$ that are in marked contrast with the Gaussian propagator of the
usual Brownian motion: (i) $P(x,t)$ generally has a non-Gaussian tail; (ii)
$P(x,t)$ reaches its maximum at a {\it non-zero} displacement.
Equation~(\ref{large x}) shows that the location of this maximum asymptotically
grows as $t^{2/3}$ when $f(D)=a$.  Thus local acceleration pushes a diffusing
particle away from the origin.


From the general expression \eqref{joint}, the marginal distribution with
respect to $D$ can also be easily obtained by integration over $x$.  We find
\begin{equation}
\label{marginalD}
P(D,t)=\Theta(D-D_0)\frac{2\sqrt{D}}{f(D)\sqrt{\pi t}}\,\,e^{-F^2/t}.
\end{equation}
In the particular case of $f(D)=a$, Eq.~\eqref{marginalD} shows that the
diffusion coefficient of the particle asymptotically grows as $t^{1/3}$.

As a byproduct, Eq.~(\ref{marginalD}) also provides the distribution of the
local time $\tau(t)$ spent by the particle in the active zone (the origin for
the present case) up to time $t$.  Using the second of Eqs.~\eqref{LE}, this
basic observable in the theory of diffusion, which has dimension of time per unit of length \cite{Majumdar:2002}, is related
to the diffusion coefficient at time $t$ by
\begin{equation}
\label{local}
\tau(t)\equiv \int_0^t\! \delta(x(t'))\,{\rm d}t' = \int_{D_0}^D\frac{{\rm d}D'}{f(D')}~.
\end{equation}
Thus the distribution of the local time, defined as $\mathcal{P}(\tau,t)$, is
given by $\mathcal{P}(\tau,t)=f(D)P(D,t)$, with $P(D,t)$ given by
Eq.~(\ref{marginalD}) and $D$ implicitly defined as a function of $\tau$ in
Eq.~(\ref{local}).  For the illustrative case of $f(D)=a$, the distribution
of the local time at time $t$ therefore is
\begin{equation}
\label{distriblocal}
\mathcal{P}(\tau,t)\!=\!\frac{2\sqrt{a\tau+D_0}}{\sqrt{\pi t}}
\exp\!\left\{-\frac{\!4\big[(a\tau\!+\!D_0)^{3/2}\!-\!D_0^{3/2}\big]^2}{9a^2t}\right\}.
\end{equation}
This result strongly contrasts with the Gaussian distribution that arises in
the case of Brownian motion, which can be recovered from
Eq.~(\ref{distriblocal}) in the limit $a\to 0$:
\begin{equation}
  \mathcal{P}_{BM}(\tau,t)=\frac{2\sqrt{D_0}}{\sqrt{\pi t}}\,\,e^{-D_0\tau^2/t}~.
\end{equation}
Notice, in particular, the typical local time for an accelerated particle
with $f(D)=a$ grows as $t^{1/3}$ instead of $t^{1/2}$ in the case of Brownian
motion.

It is worth noting an intriguing dichotomy with a discrete-time version of
local acceleration --- the ``greedy'' random walk~\cite{Ben-Naim:2004}.  In
this discrete model, the step length $\ell_k$ after the $k^{\rm th}$ return
of a random walk to the origin is given by $\ell_k=k^\alpha$.  To match with
the continuous model with $f(D)=D^\alpha$, one must choose the value $\alpha
=1/2$.  With this choice, Eq.~(13) of \cite{Ben-Naim:2004} gives, ignoring
all multiplicative factors, $P(x,t) \propto x^{1/3}/t \exp[-x^{4/3}/t]$,
which is different from (\ref{large x}).  The source of this discrepancy is
that the probability of being at the origin is not affected by the
enhancement mechanism of greedy walks \cite{Ben-Naim:2004}, while this return
probability is fundamentally modified in the case of locally-activated random
walks, as seen explicitly from the distribution of the local time
(\ref{distriblocal}).  Thus our locally activated diffusion model cannot be
viewed as the continuous limit of the greedy random walk. However, it can be shown that it is the continuous limit of a discrete space and continuous time random walk whose jump frequency is modified at each visit of the active site, which is intrinsically different from the greedy random walk. 


{\it Local Deceleration, $f(D)\!<\!0$.} Following the same analysis as that
used for local acceleration, the Laplace transform of the joint distribution
is
\begin{equation}
  \label{fneg}
  {\widehat P}=\frac{\Theta(D_0-D)}{|f(D)|}\,e^{-Z\sqrt{s}}
  -\frac{\widehat{\lambda}(s)}{s}\delta(x)\,\delta(D)\,,
\end{equation}
where $\widehat{\lambda}$ is the Laplace transform of $\lambda(t)$ defined in
Eqs.~\eqref{FP} and \eqref{def lambda}.  Using these defining relations for
$\lambda(t)$, Eq.~(\ref{fneg}) gives
\begin{equation}
\label{lambda}
\widehat{\lambda}(s)=\lim_{D\to0}\big[f(D)\,{\widehat P}(0,D,s)\big]=
-e^{-\sqrt{4s}\,\,\overline{F}}~,
\end{equation}
where we define
\begin{equation*}
\overline{F}(D)\equiv \!\int_{0}^{D_0} \frac{\sqrt{D'}}{|f(D')|}\,\,{\rm d}D'\,.
\end{equation*}
In this result for $\widehat{\lambda}$, we have used $\delta(D)f(D)=0$, since
$f(0)=0$ by the definition of our model.  The important feature of
Eq.~\eqref{lambda} is that $\widehat{\lambda}(s)=0$ as soon as $\overline{F}$
diverges.


Thus our final result is
\begin{equation}
  {\widehat P}(x,D,s)\!=\!\frac{\Theta(D_0\!-\!D)}{|f(D)|}\,e^{-Z\sqrt{s}}
  +\frac{\delta(x)\delta(D)}{s}\, e^{-\sqrt{4s}\,\,\overline{F}}~,
\end{equation}
which gives, after Laplace inversion,
\begin{align}
\label{jointdecel}
P(x,D,t)\!=\!\frac{\Theta(D\!-\!D_0)}{|f(D)|}\frac{Z\,e^{-Z^2/4t}}{\sqrt{4\pi t^3}}\,
+T(t)\delta(x)\delta(D).
\end{align}
Here 
\begin{equation}
\label{survival}
T(t)={\rm erfc}\Big(\frac{1}{\sqrt{t}}\int_{0}^{D_0} \frac{\sqrt{D'}}{|f(D')|}\,{\rm d}D'\Big)
\end{equation}
is the trapping probability, namely, the probability that the particle
becomes stuck at $x=0$ by time $t$ because the diffusion coefficient has
reached zero.  As a corollary, the survival probability is given by $S(t)=1-T(t)$, and we
have obtained this quantity for an explicitly non-Markovian process.  We also
mention that, as in the case of local acceleration, the joint distribution
easily gives the marginal distributions of the position and the diffusion
coefficient, as well as the local time.

A fundamental consequence of the local deceleration of a Brownian particle is
that two different dynamical regimes emerge.  We illustrate these regimes for
the particular case where $f(D)=-D^\beta$ as $D\to 0$.  If the deceleration
is sufficiently strong, which occurs when $\beta< 3/2$, there is a non-zero
probability for the particle to get trapped at the origin.  More precisely,
the survival probability has the asymptotic behavior
  \begin{equation}
S(t)\sim {\frac {4 D_0^{3/2-\beta}}{\sqrt {\pi t} \left( 3
-2\,\beta \right) }}\to 0\qquad\qquad t\to\infty\,.
\end{equation}
Thus in this regime of strong deceleration, the survival probability has the
same scaling with time as in the case of a usual Brownian particle in the
presence of a perfect trap.  In the opposite case of $\beta\ge 3/2$, then
$S(t)=1$ for all $t>0$ and the particle never gets trapped at the origin.
Thus a locally decelerated Brownian particle undergoes a dynamical transition
to the absorbing state $(x=0,D=0)$ as the deceleration strength increases.
Mathematically, this transition occurs at the point where $\overline{F}$ is
no longer divergent.

In conclusion, we introduced a minimal model of locally activated diffusion,
in which the diffusion coefficient of a Brownian particle is modified in a
prescribed way at each crossing of the origin.  In one dimension, a purely
diffusing particle hits the origin of the order of $\sqrt{t}$ times after a
time $t$.  Consequently, the local activation mechanism is repeatedly invoked
during the trajectory of a Brownian particle.  Thus the asymptotic dynamics
of a Brownian particle is globally affected, leading to markedly different
behavior than that of pure diffusion.  Since the unusual properties of local
activation rely on the recurrence of Brownian motion, we anticipate that
qualitatively similar, but quantitatively distinct, behavior would arise in
two dimensions.

Our model encompasses both the situations where the Brownian particle is
locally accelerated or decelerated.  For local acceleration, the probability
distribution is non-Gaussian and multi-peaked, with maxima away from the
origin no matter how weak the acceleration.  For sufficiently weak local
deceleration, a Brownian particle manages to avoid getting trapped at the
origin in spite of its recurrence.  However, for strong deceleration, there
is a dynamical transition to an absorbing state in which the particle
ultimately gets trapped at the origin.

In the context of atherosclerosis mentioned initially, the dynamical
transition to an absorbing state can be viewed as a minimal mechanism that
leads to the segregation of macrophages in lipid-enriched regions and thus to
the formation of atherosclerotic plaques.  Our model suggests that even in absence of chemical signals (such as chemokines or cytokines) that can bias the motion of cells, there
exists a critical intensity of the mobility decrease, which depends on the
local lipid concentration, beyond which an atherosclerotic plaque will occur.
Our model can also help understand the kinetics of this plaque formation.

OB is supported by the  ERC starting grant FPTOpt-277998. NM is supported by the European Project ARTreat FP7 - 224297. SR gratefully acknowledges financial support from NSF Grant No.\ DMR-0906504
and the hospitality of UPMC where this work was initiated.



\begin{thebibliography}{28}
\expandafter\ifx\csname natexlab\endcsname\relax\def\natexlab#1{#1}\fi
\expandafter\ifx\csname bibnamefont\endcsname\relax
  \def\bibnamefont#1{#1}\fi
\expandafter\ifx\csname bibfnamefont\endcsname\relax
  \def\bibfnamefont#1{#1}\fi
\expandafter\ifx\csname citenamefont\endcsname\relax
  \def\citenamefont#1{#1}\fi
\expandafter\ifx\csname url\endcsname\relax
  \def\url#1{\texttt{#1}}\fi
\expandafter\ifx\csname urlprefix\endcsname\relax\def\urlprefix{URL }\fi
\providecommand{\bibinfo}[2]{#2}
\providecommand{\eprint}[2][]{\url{#2}}

\bibitem[{\citenamefont{Toner et~al.}(2005)\citenamefont{Toner, Tu, and
  Ramaswamy}}]{Toner:2005}
\bibinfo{author}{\bibfnamefont{J.}~\bibnamefont{Toner}},
  \bibinfo{author}{\bibfnamefont{Y.}~\bibnamefont{Tu}}, \bibnamefont{and}
  \bibinfo{author}{\bibfnamefont{S.}~\bibnamefont{Ramaswamy}},
  \bibinfo{journal}{Annals of Physics} \textbf{\bibinfo{volume}{318}},
  \bibinfo{pages}{170} (\bibinfo{year}{2005}).

\bibitem[{\citenamefont{Puglisi et~al.}(1999)\citenamefont{Puglisi, Loreto,
  Marini Bettolo~Marconi, and Vulpiani}}]{PhysRevE.59.5582}
\bibinfo{author}{\bibfnamefont{A.}~\bibnamefont{Puglisi}},
  \bibinfo{author}{\bibfnamefont{V.}~\bibnamefont{Loreto}},
  \bibinfo{author}{\bibfnamefont{U.}~\bibnamefont{Marini Bettolo~Marconi}},
  \bibnamefont{and} \bibinfo{author}{\bibfnamefont{A.}~\bibnamefont{Vulpiani}},
  \bibinfo{journal}{Phys. Rev. E} \textbf{\bibinfo{volume}{59}},
  \bibinfo{pages}{5582} (\bibinfo{year}{1999}).

\bibitem[{\citenamefont{Toner and Tu}(1995)}]{PhysRevLett.75.4326}
\bibinfo{author}{\bibfnamefont{J.}~\bibnamefont{Toner}} \bibnamefont{and}
  \bibinfo{author}{\bibfnamefont{Y.}~\bibnamefont{Tu}}, \bibinfo{journal}{Phys.
  Rev. Lett.} \textbf{\bibinfo{volume}{75}}, \bibinfo{pages}{4326}
  (\bibinfo{year}{1995}).

\bibitem[{\citenamefont{Voituriez et~al.}(2005)\citenamefont{Voituriez, Joanny,
  and Prost}}]{Voituriez:2005}
\bibinfo{author}{\bibfnamefont{R.}~\bibnamefont{Voituriez}},
  \bibinfo{author}{\bibfnamefont{J.~F.} \bibnamefont{Joanny}},
  \bibnamefont{and} \bibinfo{author}{\bibfnamefont{J.}~\bibnamefont{Prost}},
  \bibinfo{journal}{Europhysics Letters} \textbf{\bibinfo{volume}{70}},
  \bibinfo{pages}{404} (\bibinfo{year}{2005}).

\bibitem[{\citenamefont{Evans and Majumdar}(2011)}]{PhysRevLett.106.160601}
\bibinfo{author}{\bibfnamefont{M.~R.} \bibnamefont{Evans}} \bibnamefont{and}
  \bibinfo{author}{\bibfnamefont{S.~N.} \bibnamefont{Majumdar}},
  \bibinfo{journal}{Phys. Rev. Lett.} \textbf{\bibinfo{volume}{106}},
  \bibinfo{pages}{160601} (\bibinfo{year}{2011}).

\bibitem[{\citenamefont{{Evans} and {Majumdar}}(2011)}]{2011arXiv1107.4225E}
\bibinfo{author}{\bibfnamefont{M.~R.} \bibnamefont{{Evans}}} \bibnamefont{and}
  \bibinfo{author}{\bibfnamefont{S.~N.} \bibnamefont{{Majumdar}}},
  \bibinfo{journal}{ArXiv e-prints}  (\bibinfo{year}{2011}),
  \eprint{1107.4225}.

\bibitem[{\citenamefont{Romanczuk and
  Schimansky-Geier}(2011)}]{PhysRevLett.106.230601}
\bibinfo{author}{\bibfnamefont{P.}~\bibnamefont{Romanczuk}} \bibnamefont{and}
  \bibinfo{author}{\bibfnamefont{L.}~\bibnamefont{Schimansky-Geier}},
  \bibinfo{journal}{Phys. Rev. Lett.} \textbf{\bibinfo{volume}{106}},
  \bibinfo{pages}{230601} (\bibinfo{year}{2011}).

\bibitem[{\citenamefont{Hughes}(1995)}]{Hughes:1995a}
\bibinfo{author}{\bibfnamefont{B.~D.} \bibnamefont{Hughes}},
  \emph{\bibinfo{title}{Random Walks and Random Environments}}
  (\bibinfo{publisher}{Oxford Press}, \bibinfo{year}{1995}).

\bibitem[{\citenamefont{Redner and Kang}(1984)}]{PhysRevA.30.3362}
\bibinfo{author}{\bibfnamefont{S.}~\bibnamefont{Redner}} \bibnamefont{and}
  \bibinfo{author}{\bibfnamefont{K.}~\bibnamefont{Kang}},
  \bibinfo{journal}{Phys. Rev. A} \textbf{\bibinfo{volume}{30}},
  \bibinfo{pages}{3362} (\bibinfo{year}{1984}).

\bibitem[{\citenamefont{Majumdar}(1990)}]{SN1990207}
\bibinfo{author}{\bibfnamefont{S.~N.} \bibnamefont{Majumdar}},
  \bibinfo{journal}{Physica A: Statistical Mechanics and its Applications}
  \textbf{\bibinfo{volume}{169}}, \bibinfo{pages}{207 } (\bibinfo{year}{1990}).

\bibitem[{\citenamefont{Weiss}(1994)}]{Weiss:1994}
\bibinfo{author}{\bibfnamefont{G.}~\bibnamefont{Weiss}},
  \emph{\bibinfo{title}{Aspects and Applications of the Random Walk}}
  (\bibinfo{publisher}{Amsterdam, Netherlands: North-Holland},
  \bibinfo{year}{1994}).

\bibitem[{\citenamefont{Tejedor et~al.}(2010)\citenamefont{Tejedor, B\'enichou,
  Voituriez, and Moreau}}]{Tejedor:2010a}
\bibinfo{author}{\bibfnamefont{V.}~\bibnamefont{Tejedor}},
  \bibinfo{author}{\bibfnamefont{O.}~\bibnamefont{B\'enichou}},
  \bibinfo{author}{\bibfnamefont{R.}~\bibnamefont{Voituriez}},
  \bibnamefont{and} \bibinfo{author}{\bibfnamefont{M.}~\bibnamefont{Moreau}},
  \bibinfo{journal}{Physical Review E} \textbf{\bibinfo{volume}{82}},
  \bibinfo{pages}{056106} (\bibinfo{year}{2010}).

\bibitem[{\citenamefont{L.}(1960)}]{L.:1960}
\bibinfo{author}{\bibfnamefont{C.~K.} \bibnamefont{L.}},
  \emph{\bibinfo{title}{Markov Chains with Stationary Transition
  Probabilities}} (\bibinfo{publisher}{Springer, Berlin},
  \bibinfo{year}{1960}).

\bibitem[{\citenamefont{Zia and Toroczkai}(1998)}]{Zia:1998}
\bibinfo{author}{\bibfnamefont{R.~K.~P.} \bibnamefont{Zia}} \bibnamefont{and}
  \bibinfo{author}{\bibfnamefont{Z.}~\bibnamefont{Toroczkai}},
  \bibinfo{journal}{Journal of Physics A: Mathematical and General}
  \textbf{\bibinfo{volume}{31}}, \bibinfo{pages}{9667} (\bibinfo{year}{1998}).

\bibitem[{\citenamefont{Benichou and Oshanin}(2002)}]{Benichou:2002qq}
\bibinfo{author}{\bibfnamefont{O.}~\bibnamefont{B\'enichou}} \bibnamefont{and}
  \bibinfo{author}{\bibfnamefont{G.}~\bibnamefont{Oshanin}},
  \bibinfo{journal}{Phys Rev E Stat Nonlin Soft Matter Phys}
  \textbf{\bibinfo{volume}{66}}, \bibinfo{pages}{031101}
  (\bibinfo{year}{2002}).

\bibitem[{\citenamefont{Calvez et~al.}(2008)\citenamefont{Calvez, Ebde,
  Meunier, Raoult, and Ruznakova}}]{Calvez:2008}
\bibinfo{author}{\bibfnamefont{V.}~\bibnamefont{Calvez}},
  \bibinfo{author}{\bibfnamefont{A.}~\bibnamefont{Ebde}},
  \bibinfo{author}{\bibfnamefont{N.}~\bibnamefont{Meunier}},
  \bibinfo{author}{\bibfnamefont{A.}~\bibnamefont{Raoult}}, \bibnamefont{and}
  \bibinfo{author}{\bibfnamefont{G.}~\bibnamefont{Ruznakova}},
  \bibinfo{journal}{ESAIM Proc.} \textbf{\bibinfo{volume}{28}},
  \bibinfo{pages}{1} (\bibinfo{year}{2008}).

\bibitem[{\citenamefont{Calvez et~al.}(2009)\citenamefont{Calvez, Houot,
  Meunier, Raoult, and Ruznakova}}]{Calvez:2009}
\bibinfo{author}{\bibfnamefont{V.}~\bibnamefont{Calvez}},
  \bibinfo{author}{\bibfnamefont{J.}~\bibnamefont{Houot}},
  \bibinfo{author}{\bibfnamefont{N.}~\bibnamefont{Meunier}},
  \bibinfo{author}{\bibfnamefont{A.}~\bibnamefont{Raoult}}, \bibnamefont{and}
  \bibinfo{author}{\bibfnamefont{G.}~\bibnamefont{Ruznakova}},
  \bibinfo{journal}{ESAIM Proc.} \textbf{\bibinfo{volume}{30}},
  \bibinfo{pages}{1} (\bibinfo{year}{2009}).

\bibitem[{\citenamefont{de~Winther and Hofker}(2000)}]{Winther:2000}
\bibinfo{author}{\bibfnamefont{M.~P.~J.} \bibnamefont{de~Winther}}
  \bibnamefont{and} \bibinfo{author}{\bibfnamefont{M.~H.}
  \bibnamefont{Hofker}}, \bibinfo{journal}{The Journal of Clinical
  Investigation} \textbf{\bibinfo{volume}{105}}, \bibinfo{pages}{1039}
  (\bibinfo{year}{2000}).

\bibitem[{\citenamefont{Caro}(2009)}]{Caro:2009}
\bibinfo{author}{\bibfnamefont{C.~G.} \bibnamefont{Caro}},
  \bibinfo{journal}{Arterioscler. Thromb. Vasc. Biol.}
  \textbf{\bibinfo{volume}{29}}, \bibinfo{pages}{158} (\bibinfo{year}{2009}).

\bibitem[{\citenamefont{Yvan-Charvet et~al.}(2007)\citenamefont{Yvan-Charvet,
  Ranaletta, Wang, Han, Terasaka, Li, Welch, and Tall}}]{Yvan-Charvet:2007}
\bibinfo{author}{\bibfnamefont{L.}~\bibnamefont{Yvan-Charvet}},
  \bibinfo{author}{\bibfnamefont{M.}~\bibnamefont{Ranaletta}},
  \bibinfo{author}{\bibfnamefont{N.}~\bibnamefont{Wang}},
  \bibinfo{author}{\bibfnamefont{S.}~\bibnamefont{Han}},
  \bibinfo{author}{\bibfnamefont{N.}~\bibnamefont{Terasaka}},
  \bibinfo{author}{\bibfnamefont{R.}~\bibnamefont{Li}},
  \bibinfo{author}{\bibfnamefont{C.}~\bibnamefont{Welch}}, \bibnamefont{and}
  \bibinfo{author}{\bibfnamefont{A.}~\bibnamefont{Tall}}, \bibinfo{journal}{The
  Journal of Clinical Investigation} \textbf{\bibinfo{volume}{117}},
  \bibinfo{pages}{3900} (\bibinfo{year}{2007}).

\bibitem[{\citenamefont{Siegel-Axel et~al.}(2008)\citenamefont{Siegel-Axel,
  Daub, Seizer, Lindemann, and Gawaz}}]{Siegel-Axel:2008}
\bibinfo{author}{\bibfnamefont{D.}~\bibnamefont{Siegel-Axel}},
  \bibinfo{author}{\bibfnamefont{K.}~\bibnamefont{Daub}},
  \bibinfo{author}{\bibfnamefont{P.}~\bibnamefont{Seizer}},
  \bibinfo{author}{\bibfnamefont{S.}~\bibnamefont{Lindemann}},
  \bibnamefont{and} \bibinfo{author}{\bibfnamefont{M.}~\bibnamefont{Gawaz}},
  \bibinfo{journal}{Cardiovascular Research} \textbf{\bibinfo{volume}{78}},
  \bibinfo{pages}{8} (\bibinfo{year}{2008}).

\bibitem[{\citenamefont{Kampen}(1992)}]{Kampen:1992a}
\bibinfo{author}{\bibfnamefont{N.~V.} \bibnamefont{Kampen}},
  \emph{\bibinfo{title}{Stochastic Processes in Physics and Chemistry}}
  (\bibinfo{publisher}{North -Holland}, \bibinfo{year}{1992}).

\bibitem[{\citenamefont{Majumdar and Sire}(1996)}]{PhysRevLett.77.1420}
\bibinfo{author}{\bibfnamefont{S.~N.} \bibnamefont{Majumdar}} \bibnamefont{and}
  \bibinfo{author}{\bibfnamefont{C.}~\bibnamefont{Sire}},
  \bibinfo{journal}{Phys. Rev. Lett.} \textbf{\bibinfo{volume}{77}},
  \bibinfo{pages}{1420} (\bibinfo{year}{1996}).

\bibitem[{\citenamefont{Majumdar}(1999)}]{Majumdar:1999a}
\bibinfo{author}{\bibfnamefont{S.~N.} \bibnamefont{Majumdar}},
  \bibinfo{journal}{Curr. Sci.} \textbf{\bibinfo{volume}{77}},
  \bibinfo{pages}{370} (\bibinfo{year}{1999}).

\bibitem[{\citenamefont{Redner}(2001)}]{Redner:2001a}
\bibinfo{author}{\bibfnamefont{S.}~\bibnamefont{Redner}},
  \emph{\bibinfo{title}{A Guide to First-Passage Processes}}
  (\bibinfo{publisher}{Cambridge University Press, Cambridge, England},
  \bibinfo{year}{2001}).

\bibitem[{\citenamefont{Gardiner}(2004)}]{Gardiner:2004}
\bibinfo{author}{\bibfnamefont{C.}~\bibnamefont{Gardiner}},
  \emph{\bibinfo{title}{Handbook of Stochastic Methods for Physics, Chemistry
  and Natural Sciences}} (\bibinfo{publisher}{Springer}, \bibinfo{year}{2004}).

\bibitem[{\citenamefont{Majumdar and Comtet}(2002)}]{Majumdar:2002}
\bibinfo{author}{\bibfnamefont{S.~N.} \bibnamefont{Majumdar}} \bibnamefont{and}
  \bibinfo{author}{\bibfnamefont{A.}~\bibnamefont{Comtet}},
  \bibinfo{journal}{Phys. Rev. Lett.} \textbf{\bibinfo{volume}{89}},
  \bibinfo{pages}{60601} (\bibinfo{year}{2002}).

\bibitem[{\citenamefont{Ben-Naim and Redner}(2004)}]{Ben-Naim:2004}
\bibinfo{author}{\bibfnamefont{E.}~\bibnamefont{Ben-Naim}} \bibnamefont{and}
  \bibinfo{author}{\bibfnamefont{S.}~\bibnamefont{Redner}},
  \bibinfo{journal}{Journal of Physics A: Mathematical and General}
  \textbf{\bibinfo{volume}{37}}, \bibinfo{pages}{11321} (\bibinfo{year}{2004}).

\end{thebibliography}

\end{document}